\documentclass[graybox]{svmult}
\usepackage{mathptmx}       
\usepackage{helvet}         
\usepackage{courier}        
\usepackage{type1cm}        
\usepackage{epsfig,amsmath,amssymb,graphics} 
\usepackage{makeidx}         
\usepackage{graphicx}        
\usepackage{multicol}        
\usepackage[bottom]{footmisc}
\makeindex             

\begin{document}

\title*{Evolution of Systems with Power-Law Memory: Do We Have to Die?}
\author{M. Edelman}
\institute{M. Edelman \at Department of Physics, Stern College at 
Yeshiva University\\ 
245 Lexington Ave, New York, NY 10016, USA; \\
Courant Institute of Mathematical Sciences, New York University \\
251 Mercer St., New York, NY 10012, USA; \\ 
Department of Mathematics, BCC, CUNY, \\
2155 University Avenue, Bronx, New York 10453, USA,\\
 \email{edelman@cims.nyu.edu}}

\maketitle

\abstract{Various features of the development of individual living
  species, including individual humans, are programmed. Is death also
  programmed, and if yes, how is it implemented and what can be the 
underlying mechanism providing the inevitability of death? The 
hypothesis presented in this paper is based on the similarity 
of the human evolution to the evolution of simple discrete nonlinear 
fractional (with power-law memory) systems. Caputo fractional/fractional 
difference logistic map is a simple discrete system with
power-/asymptotically power-law memory and quadratic nonlinearity. In the 
area of parameters where the fixed point is unstable, its evolution starts 
as the evolution of a system with a stable fixed point but then this fixed 
point becomes unstable, suddenly breaks, and turns into a period two
point. Considered under various types of random perturbations, the time
spans of the evolution as a fixed point before the break (lifespans) 
obey the Gompertz-Makeham law, which is the observed distribution of 
the lifespans of live species, including humans. The underlying reason 
for modeling the evolution of humans by fractional systems are the 
observed power law in human memory and the viscoelastic nature of 
organ tissues of living species. Models with power-law memory may explain 
the observed decrease at very large ages of the rate of increase of the 
force of mortality and they imply limited lifespans.	
}

\section{Introduction}
\label{sec:1}

This paper is meant for publication in a collection of papers in memory 
of Valentin Afraimovich, an outstanding mathematician and a wonderful
man. It was a pleasure to be acquainted with Valentin for more than 20
years, to have conversations about life and science. The spectrum 
of his scientific interests was very wide. Although he didn't work 
in the field of fractional dynamics, he attended many talks. He told 
me once that he would work in discrete fractional calculus if 
fractional systems were considered as an infinite-dimensional 
system. Unfortunately, this approach to fractional dynamics was 
not implemented during Valentin's life.

I developed an interest in fractional calculus while working with 
George Zaslavsky at NYU on transport in Hamiltonian systems and 
billiards. Work on fractional maps introduced in \cite{T1} to 
investigate general properties of fractional (with power-law memory) 
systems was proposed by George and discrete fractional dynamics has 
been the main topic of my research during the last ten years.  

Fractional calculus has many applications, which include 
(the number of publications on applications of fractional 
calculus is overwhelming and the references here are just examples) 
transport (anomalous diffusion) in Hamiltonian systems and 
billiards \cite{ZasBook}, systems with long-range 
interactions \cite{LR1,LR2}, electromagnetic fields in 
dielectric media \cite{TDia2008a,TDia2009,T2}, viscoelastic 
materials and rheology \cite{Visc1,Visc2,Visc3}, fractional 
control \cite{FrCon}, and many others.

\section{Fractional Dynamical Systems}
\label{sec:2}

Fractional differences/derivatives are convolutions with power-law 
functions. Therefore, fractional space derivatives appear in 
equations of distributed in space systems, but the real fractional 
dynamics is the dynamics of systems with power-law memory. In such 
systems, which are called fractional dynamical systems, new values 
of the system's variables depend on the whole history of the 
system's evolution. In what follows we will use the fractional 
Caputo derivative defined as (see e.g. \cite{KST})
{\setlength\arraycolsep{0.5pt}
\begin{equation}
_0^CD^{\alpha}_t x(t)=_0I^{n-\alpha}_t \ D^n_t x(t)
=\frac{1}{\Gamma(n-\alpha)}  \int^{t}_0 
\frac{ D^n_{\tau}x(\tau) d \tau}{(t-\tau)^{\alpha-n+1}}  \quad (n-1 <\alpha \le n),
\label{Cap}
\end{equation}
}
where $ D^n_t x(t)$ is a regular derivative of the $n$th order.
We will also use 
examples based on the Caputo fractional/fractional difference standard 
and logistic maps. They are particular forms of the 
fractional/fractional difference universal map (see \cite{ME8,ME9,Brazil,HBV2}).
The m-dimensional Caputo fractional universal map can be written as 
\begin{equation}
x^{(s)}_{n+1}= \sum^{m-s-1}_{k=0}\frac{x^{(k+s)}_0}{k!}h^k(n+1)^{k} 
-\frac{h^{\alpha-s}}{\Gamma(\alpha-s)}\sum^{n}_{k=0} G_K(x_k) (n-k+1)^{\alpha-s-1},
\label{FrCM}
\end{equation}
where $\alpha \in \mathbb{R}$, $\alpha\ge 0$, $m=\lceil \alpha \rceil$,
$n \in \mathbb{Z}$, 
$n \ge 0$, $s=0,1,...,m-1$. In this paper we assume that $x^{(1)}$ is a 
momentum $p$ and in all maps (including fractional difference maps) consider the case $h=1$.

The m-dimensional Caputo fractional difference universal map can be written as 
\begin{eqnarray} 
&&x_{n+1} =   \sum^{m-1}_{k=0}\frac{\Delta^{k}x(0)}{k!}(n+1)^{(k)} 
\nonumber \\
&&-\frac{1}{\Gamma(\alpha)}  
\sum^{n+1-m}_{s=0}(n-s-m+\alpha)^{(\alpha-1)} 
G_K(x_{s+m-1}),
\label{FalFacMap}
\end{eqnarray}
where $\alpha \in \mathbb{R}$, $\alpha\ge 0$, $m=\lceil \alpha \rceil$,
$n \in \mathbb{Z}$, 
$n \ge 0$, $s=0,1,...,m-1$,
\begin{equation}
\Delta^{k} x(0) = c_k, \ \ \ k=0, 1, ..., m-1, \ \ \ 
m=\lceil \alpha \rceil
\label{LemmaDifICn}
\end{equation}
are the initial conditions, and the falling factorial $t^{(\alpha)}$ is
defined as
\begin{equation}
t^{(\alpha)} =\frac{\Gamma(t+1)}{\Gamma(t+1-\alpha)}, \ \ t\ne -1, -2, -3....
\label{FrFac}
\end{equation}

In the case $\alpha=1$ maps Eq.~(\ref{FrCM}) and 
Eq.~(\ref{FalFacMap}) converge to the regular logistic map
\begin{equation}
x_{n+1}= Kx_{n}(1-x_{n})
\label{LogEq}
\end{equation} 
if (see \cite{ME3,ME4,ME7,Brazil,Chaos2018,HBV2} for 
fractional maps and \cite{ME9,Brazil,Chaos2018,HBV2} for fractional  difference maps)
\begin{equation}
G_K(x)= x-Kx(1-x)
\label{LogEqG}.
\end{equation} 
These maps are called Caputo fractional (in the case 
of Eq.~(\ref{FrCM})) or fractional difference (in the 
case of Eq.~(\ref{FalFacMap})) logistic $\alpha$-families 
of maps (or simply logistic Caputo fractional/fractional difference maps).

In the case $\alpha=2$ maps Eq.~(\ref{FrCM}) and 
Eq.~(\ref{FalFacMap}) converge to the regular standard map (see \cite{Chirikov})
\begin{equation}
p_{n+1}= p_{n} - K \sin(x_n) \ \ \ ({\rm mod} \ 2\pi ), 
\label{SMp}
\end{equation}
\begin{equation}
x_{n+1}= x_{n}+ p_{n+1} \ \ \ ({\rm mod} \ 2\pi )
\label{SMx}
\end{equation}
if (see \cite{ME3,ME4,ME7,Brazil,Chaos2018,HBV2,T2009a,T2009b,T2} 
for fractional maps and \cite{ME8,ME9,Brazil,Chaos2018,HBV2} for 
fractional  difference maps)
\begin{equation}
G_K(x)=K \sin(x).
\label{SFM}
\end{equation}
These maps are called Caputo fractional (in the case 
of Eq.~(\ref{FrCM})) or fractional difference (in the 
case of Eq.~(\ref{FalFacMap})) standard $\alpha$-families of 
maps (or simply standard Caputo fractional/fractional difference maps).

We'll use the fractional Riemann-Liouville standard map for 
$1< \alpha \le 2$ in the form 
(see \cite{Brazil,Chaos2018,HBV2,ME1,T2,T1})
\begin{equation} 
\label{2}
p_{n+1} = p_n - K \sin x_n ,
\end{equation}
$$
x_{n+1} = \frac{1}{\Gamma (\alpha )} 
\sum_{i=0}^{n} p_{i+1}V_{\alpha}(n-i+1) , \ \ \ \ ({\rm mod} \ 2\pi ) ,
$$
where 
\begin{equation} 
\label{4}
V_{\alpha}(m)=m^{\alpha -1}-(m-1)^{\alpha -1}. 
\end{equation} 

\begin{figure}[!t]
\begin{center}
\includegraphics[width=0.9\textwidth]{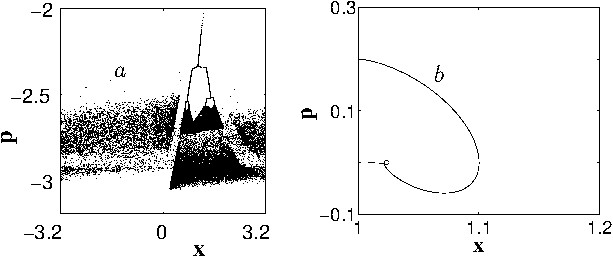}
\vspace{-0.25cm}
\caption{ 
(a) 20000 iterations on each of two overlapping attractors 
of the Caputo fractional standard map's phase space with $K=4.5$, 
$\alpha=1.02$, and $x_0=0$; the cascade of bifurcations type 
trajectory has initial momentum $p_0=-1.8855$ and the chaotic attractor
$p_0=-2.5135$. 
(b) A self-intersecting phase space trajectory of the 
fractional Caputo Duffing equation 
$_0^CD^{1.5}_t x(t)=x(1-x^2)$, $t\in [0,40]$ 
with the initial conditions $x(0)=1$
and $dx/dt(0)=0.199$. Fractional Caputo derivative is defined by 
Eq.~(\ref{Cap}).
}
\end{center}
\label{fig1}
\end{figure}

Properties of fractional dynamical systems are different f
rom properties of the regular dynamical systems in the following:
\begin{itemize}
\item
{
Chaotic attractors in fractional dynamical systems may overlap and 
trajectories may intersect in continuous fractional systems of low orders
Fig.~1. This leads to the inapplicability of the  Poincar$\acute{e}$-Bendixson 
theorem to fractional systems. Therefore, in continuous fractional
systems of low orders non-existence of chaos is only a
conjecture (see \cite{Deshpande,Chaos2015}). 
}
\item
{ 
In fractional systems periodic sinks, except fixed points, 
may exist only in an asymptotic sense. Asymptotically 
attracting points may not belong to their own basins of attraction 
(see \cite{ME2,ME3,ME4}). A trajectory starting from an asymptotically 
attracting point may end attracted to a different asymptotically attracting point.  
The rate at which a trajectory is approaching an attracting point depends 
on initial conditions. Trajectories originating from a basin of attraction may 
converge faster (as $x_n \approx n^{-1-\alpha}$ for the fractional 
Riemann-Liuoville standard map, see Fig.~1 from \cite{ME3}) than
trajectories originating from the chaotic sea (as $x_n \approx n^{-\alpha}$). Fractional differential/difference equations do not have periodic solutions except fixed points; they may have asymptotically periodic solutions instead (see, e.g., \cite{PerC1,PerD1,PerD2,PerC2,PerC3,PerC4,PerC5}).
}
\item
{A new type of attractors, nonexistent in regular systems, cascade of bifurcations type trajectories (CBTT), 
Fig.~2 and Fig.~3, 
is a general feature of fractional systems. A CBTT may start converging to a period $2^n$ sink, but then bifurcate and start converging to a period $2^{n+1}$ sink, and so on. CBTT may end their evolution as a period $2^{n+m}$ sink or in chaos \cite{ME2,HBV4,ME5}. 
}
\item
{Integer members of fractional $\alpha$-families of maps are volume preserving \cite{ME4,ME8}. Fractional members are not volume preserving - they behave like the member of the same family of the higher integer order with dissipation \cite{ZSE}. Types of attractors which may 
exist in fractional systems include sinks, limiting cycles, and chaotic 
attractors \cite{AttrC1,ME3,ME7,ME5,AttrC2}.}
\item{Dependence of fractional/fractional difference $\alpha$-families of maps on the nonlinearity parameter $K$ and the memory parameter (order of a fractional system) $\alpha$ can be described by two-dimensional bifurcation diagrams 
Fig.~4. 
This, in turn, leads to the dependence of the map's bifurcations on the memory parameter. Examples of $x$-$\alpha$ bifurcation diagrams are presented in 
Fig.~5. 
This allows an additional way of controlling fractional systems by manipulating their orders $\alpha$. 
}
\end{itemize}
\begin{figure}[!t]
\begin{center}
\includegraphics[width=0.9\textwidth]{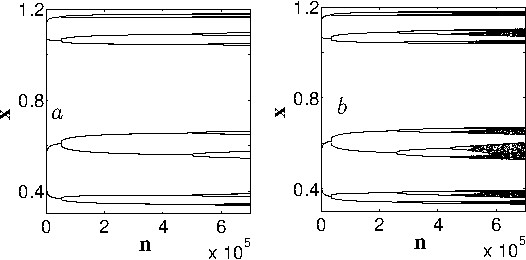}
\vspace{-0.25cm}
\caption{ 
Cascade of bifurcations type trajectories (CBTT) in the
Caputo logistic fractional maps, Eq.(\ref{FrCM}) with $h=1$ and $G_K(x)=x-Kx(1-x)$. In both cases $\alpha=0.1$ and $x_0=0.001$: 
(a) in the case $K=22.37$, the last bifurcation from the period  
$T=8$ to the period $T=16$ occurs after approximately $5 \times 10^5$ iterations; 
(b) in the case $K=22.423$, after approximately 
$5 \times 10^5$ iterations the trajectory becomes chaotic.
}
\end{center}
\label{fig2}
\end{figure}
\begin{figure}[!t]
\begin{center}
\includegraphics[width=0.9\textwidth]{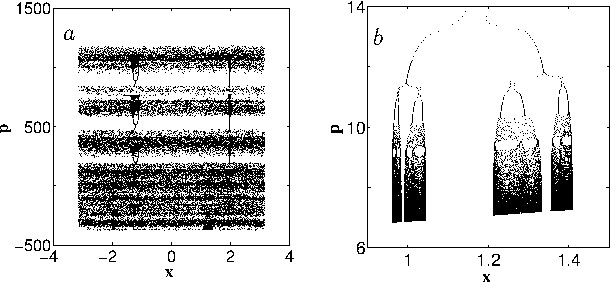}
\vspace{-0.25cm}
\caption{  
CBTT in the Riemann-Liouville fractional standard map with $1<\alpha<2$, Eq.~(\ref{2}): 
(a) 120000 iterations on a single trajectory with $K=4.5$, 
$\alpha=1.65$, and $p_0=0.3$. The trajectory occasionally sticks to one of the CBTTs (intermittent CBTT) but then always recovers into the chaotic sea. 
(b) 100000 iterations on a trajectory with $K=3.5$, 
$\alpha=1.1$, and $p_0=20$. The trajectory very fast turns into a CBTT
which slowly converges to a fractal type set.
}
\end{center}
\label{fig3}
\end{figure}
\begin{figure}[!t]
\begin{center}
\includegraphics[width=0.9\textwidth]{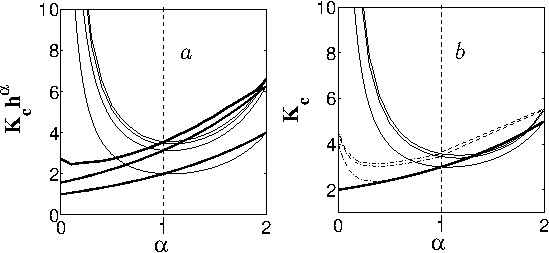}
\vspace{-0.25cm}
\caption{  
2D bifurcation diagrams for the fractional (solid thing lines) and
fractional difference (bold and dashed lines) Caputo standard (a) 
and $h=1$ logistic (b) maps. The first bifurcation, transition from a stable fixed point to a stable period two ($T=2$) sink, 
occurs on the bottom curves.
A $T=2$ sink (in the case of the standard $\alpha$-families of maps
an antisymmetric T=2 sink with $x_{n+1}=-x_n$)
is stable between the bottom and the middle curves. Transition 
to chaos occurs on the top curves. For the standard fractional map transition from a $T=2$ to $T=4$ sink
occurs on the line below the top line (the third from the bottom line). 
Period doubling bifurcations leading to
chaos occur in the narrow band between the two top curves.
This figure is reprinted from \cite{Chaos2018}, with the permission of AIP Publishing.   
}
\end{center}
\label{fig4}
\end{figure}
\begin{figure}[!t]
\begin{center}
\includegraphics[width=0.9\textwidth]{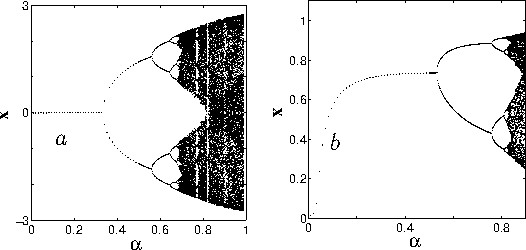}
\vspace{-0.25cm}
\caption{ The memory $\alpha$-$x$ bifurcation diagrams for fractional Caputo standard (a) and logistic (b) maps obtained after 5000 iterations.
$K=4.2$ in (a) and $K=3.8$ in (b).   
}
\end{center}
\label{fig5}
\end{figure}

\section{Power-Law Memory (Fractionality) in Biological Systems}
\label{sec:3}

The following review of studies of memory in biological systems emphasizes the role of nonlinear fractional (with power-law memory) dynamics described by the fractional differential/difference equations of the orders $0<\alpha<2$, and especially, $\alpha$ close to zero. 

Memory, as a significant property of humans, is a subject of extensive biophysical and psychological research. The power-law forgetting, the decay of the accuracy on memory tasks as $\sim t^{-\beta}$, with $0<\beta<1$, has been demonstrated in experiments described in  
\cite{Kahana,Rubin,Wixted1,Wixted2,Adaptation1}. 
Let us note here that fractional maps of the orders $0<\alpha<1$ are maps with power-law, $-\beta=\alpha-1$, decaying memory, where $0<\beta<1$ \cite{ME3}. 
Human learning is also characterized by power-law memory: the reduction in reaction times that comes with
practice is a power function of the number of training trials \cite{Anderson}.
Dynamics of biological systems at levels ranging from single ion channels up to human psychophysics in \cite{Adaptation3,Adaptation4,Adaptation2,Adaptation5,Adaptation1,Adaptation6} is described by application of power-law adaptation.

The underlying reasons for human's power-law memory can be related to the power-law memory of its building blocks, from individual neurons and proteins to tissues of individual organs. 
Processing of the external stimuli by individual neurons, as it has been shown in \cite{Neuron3,Neuron4,Neuron5}, can be described by fractional differentiation with the orders of derivatives $\alpha \in [0,1]$. 
For example, in the case of neocortical pyramidal neurons this order is $\alpha \approx 0.15$.
The power-law memory kernel with the
exponent $-0.51 \pm 0.07$ is demonstrated in fluctuations within 
single protein molecules (see \cite{Protein}).
 
As mentioned at the end of Section~\ref{sec:1}, viscoelasticity is one of the most important applications of fractional calculus. Viscoelastic materials act as substances with power-law memory and their behavior can be described by the fractional differential equations. It was demonstrated in many publications that human tissues are viscoelastic,
which is related to the viscoelasticity of the components of tissues: 
structural proteins, cells, extracellular matrices, and so on. 
Earlier references (prior to 2014) related to various organs may be found in \cite{ME7}. For more recent results on viscoelasticity of the brain tissue see \cite{TissueBrain3}, cardiovascular tissues \cite{CardVasc}, human tracheal tissues \cite{Trac}, human skin \cite{Skin}, human bladder tumours \cite{BTum}, human vocal fold tissues \cite{Voc}, and many other publications.

Additional areas of application of fractional calculus in biology include
fractional wave-propagation in biological tissues \cite{TissueWaves1,TissueWaves2,TissueWaves3,TissueWaves4}, 
bioengineering (bioelecrodes, biomechanics, and bioimaging) \cite{Magin}, 
population biology and epidemiology \cite{PopBioBook2001,HoppBook1975}).

\section{Inevitability of death}
\label{sec:4}

Humans live and die. There is no way to stop aging. According to the Guinness World Records and the Gerontology Research Group, the longest  
living person whose dates of birth and death were verified (although not without controversies), Jeanne Calment (1875--1997), lived to 122. Recent deaths of people with whom I had close relationships, George Zaslavsky, Valentin Afraimovich, and related to them Vladimir Arnold, stimulated me to write this note. 
All of them lived for approximately 73 years and died, when they had active scientific plans and close family relationships. Tragic, untimely, unjustifiable deaths. And this is the reality: however smart, intelligent, and athletic we are, however healthy our diet is, we all die. There are some religious, philosophical, and scientific thoughts about this inevitability of death.      

The religious explanation depends on which religion. In Christianity,
Adam and Eve, the first humans, lost their lives because they sinned against God. (Genesis 3:17-19). The Bible says: ``Through one man sin entered into the world and death through sin, and thus death spread to all men because they had all sinned.'' (Romans 5:12). 
In Buddhism and Hinduism death is considered as a natural part of the cycle of life followed by rebirth.
In Judaism death is not a tragedy - it is a natural process.
Deaths, like lives, have meaning and are all part of G-d's plan.
Islam believes in soul's continuous existence with transformed physical existence and a Day of Judgment deciding human beings' eternal destination to Paradise or Hell.

Life and death are topics of many philosophical discussions. Evolutionary aspects can be found, for example, in the views of Heraclitus, Nietzsche, and Hegel. Life and death are two opposites which are inseparable, and Heraclitus apparently believed in the cyclic recurrence of all things, including our lives. In Hegel's philosophy the recurrence is not cyclic but spiral.

\subsection{Evolution and Lifespan}
\label{sec:4.1}

In 1889 August Weismann, in his interpretation and formulation of the mechanism of Darwinian evolution, theorized that aging is a part of life's program because the old need to be removed to make room for the next generation to sustain the evolution \cite{Weismann}.

To list the main modern approaches in the evolutionary biology we'll follow (very briefly) Joshua Mitteldorf's review \cite{Mitteldorf}. According to Mitteldorf, the four main theoretical explanations of aging are: 
\begin{itemize}
\item
{
{\it Aging as the accumulation of damage}. This approach is supported by the similarities between mortality curves and graphs of the failure rates in the non-living world (see, e.g., examples related to the Gompertz law on page 56 of the Gavrilovs' book \cite{GavGavBook}). Corresponding mathematical models can be found in chapter six of the book (see also \cite{GavGavArt}). This approach is lacking the natural selection argumentation and the only way to slow aging here is to prevent or repair the accumulated random damage. Arguments against this approach include the difference in the interaction with the environment between living and non-living matter (possibility of the energy accumulation from the environment and dumping an excess of entropy into the surroundings), difference in the repair mechanisms (internal somatic repairs are quite different from the external repair of machines), and here we would add the presence of memory as a significant feature of living matter. 
}
\item
{
{\it Aging as irrelevant to evolution}. The original theoretical idea of Medawar \cite{Medawar} supported by Edney and Gill \cite{EdneyGill} is based on the fact that in the case of the low selection pressure  mutational load alone could explain the evolutionary emergence of aging. The corresponding theory is called Mutation Accumulation (MA). Demographic surveys \cite{Ricklefs} and the existence of a conserved genetic basis for aging \cite{GuarenteKenyon,Kenyon} contradict the MA theory. An additional argument against the MA theory is low genetic variance, which is the proper measure of selection pressure, and the long-time existence of aging controlling families of genes \cite{Fisher,Promislow,Tatar}. 

At the cellular level, apoptosis and telomeric aging \cite{Clark,Gordeeva} are two well-known mechanisms of programmed cellular death. They have always been assumed to be beneficial for the individual adaptation.
}
\item
{  
{\it Aging as the result of evolutionary tradeoffs}.
As Mittledorf mentioned, the basis for the evolutionary tradeoff theories is an "inescapable" (and this is the key word here) "tradeoff between preservation of the soma and other tasks essential to fitness, such as metabolism and reproduction". There is a significant amount of empirical data against the inescapability of the tradeoff. 

The two main branches of the pleiotropic theories are the Antagonistic Pleiotropy (AP) theory (genetic tradeoff, originated from Williams' work \cite{Williams}) and the Disposable Soma theory (tradeoff in resource allocation, originated from the work of Kirkwood \cite{Kirkwood}). 
}
\item
{
{\it Aging as a genetic program}. This approach implies that group selection to benefit the whole population is strong compared to individual adaptation based on the concept of individual competition. It contradicts modern genetic theories but has a strong support in empirical observations. For more details see \cite{Mitteldorf}. If the genetic program theory works, then one may expect that manipulations of this program may result in a significant extension of the life span. 
}
\end{itemize}
The detailed arguments in support and against of each approach can be found in Mitteldorf's paper \cite{Mitteldorf} and Mitteldorf himself is a strong proponent of the last (demographic) from the four listed approaches to the problem of the evolution of aging.

\subsection{Gompertz-Makeham Distribution}
\label{sec:4.2}

Any mathematical model of aging should satisfy the observed distribution of the lifespans. This observed distribution is the Gompertz-Makeham law
\begin{equation}
\mu(t)=A+R \exp (\gamma t),
\label{Gompertz}
\end{equation}
where $A$, $R$, and $\gamma$ are constants and $\mu$ is the force of mortality defined by the formula 
\begin{equation}
\mu(t)=-\frac{d \ln l(t)}{d t},
\label{FM}
\end{equation}
where $t$ is the age and $l(t)$ is the number of survivors from the initial population at age $t$. 
The rate of increase of the force of mortality 
is decreasing at the extremely old ages \cite{GavGavBook}.

Various models described in Ch. 6 of Gavrilivs' book \cite{GavGavBook} are based on the statistics of extremes. A heterogeneous population consists of organisms considered to be combinations of (a very large number of) subsystems with many redundant elements under stochastic perturbations. Failure of any subsystem may (in the case of a series connection) lead to the death. Therefore, the authors of the models are looking for the limiting distributions of the least values of the lifespans of a large number of subsystems. In many cases the authors of models recover the Gompertz-Makeham law. The distribution of the organisms' lifespans depends on the distribution of lifespans of its constituting elements. The Gompertz law for the subsystems' lifespans generates the Gompertz law for the lifespans of the organisms.

The models of this type with the upper limit on the organism's lifespan result in the increasing with age rate of increase of the force of mortality and contradict the observational data. The authors conclude that there can be no upper limit on the organism's lifespan.

\subsection{Fractional Dynamics Approach}
\label{sec:4.3}

As we already mentioned in the previous section, no considered models 
take into account the fact that all live species possess memory. 
As we mentioned in Sec.~\ref{sec:3}, memory is an important property of
all animals' (and humans') organ tissues (subsystems), which are viscoelastic.  

Antagonistic pleiotropy is one of the theories explaining aging. In short, it states: the better we fit, the sooner we die. Species compete to make more and better children and individuals exhaust themselves in this competition. Considering humans, the competition also includes the competition in intellects, in making discoveries (which, who knows, could be the purpose the human existence in the Universe). If the 
fittest one is not the one who succeeds in discoveries, then the
evolutionary outcome could be the extinction of the human curiosity. But
this is not what is observed. Making discoveries, thinking, requires a lot of concentration and use of organism's resources. In addition to remembering and retrieving from the brain all facts that we've learned, we keep in mind all previous steps in our reasoning. But our memory decays (or strengthens) as a power law. This could be an additional argument to consider when modeling a human being (including the evolution of aging). 

Therefore, the author believes that the use of the fractional (with power-law memory) models is a natural and appropriate approach to model a human individual. It is not trivial but possible to model a living species by a single variable and, in any case, this model will be a rough approximation of the reality. One of the ways is to consider a variable related (but not equivalent) to fitness understood, for example, as the total (potential) number of descendants produced by a certain age. 
This variable is less than one up to the age at which an organism is first capable of sexual reproduction of offspring (puberty), and then it gradually increases approaching some constant value until the organism dies (which could be long after the end of the ability to reproduce). Solutions of simple nonlinear discrete fractional equations bear a remarkable similarity to the evolution of this variable presented above. 
This similarity supports the author's proposition to use fractional models to describe the evolution of aging.

\begin{figure}[!t]
\begin{center}
\includegraphics[width=0.9\textwidth]{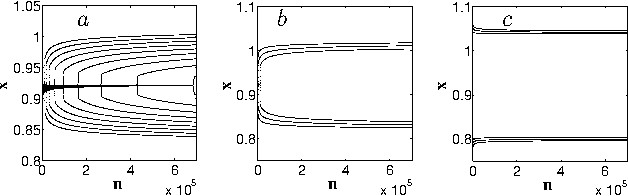}
\end{center}
\caption{Asymptotically period two trajectories for the Caputo logistic
$\alpha$-family of maps Eqs.~(\ref{FrCM})~and~(\ref{LogEqG}) 
with $\alpha=0.1$ and $K=15.5$: (a) nine
 trajectories with the initial conditions $x_0=0.29+0.04i$, $i=0,1,...,8$  
($i=0$ corresponds to the rightmost bifurcation);
(b) $x_0=0.61+0.06i$, $i=1,2,3$; (c) $x_0=0.95+0.04i$, $i=1,2,3$. As $ n
\rightarrow \infty $ all trajectories converge to the limiting values
$x_{lim1}=0.80629$ and  $x_{lim2}=1.036030$. 
The unstable fixed point is $x_{lim0}=(K-1)/K=0.93548$.
This figure is reprinted from \cite{Chaos2018}, 
with the permission of AIP Publishing.
}
\label{fig6}
\end{figure}
\begin{figure}[htb]
\centering
  \begin{tabular}{@{}cccc@{}}
    \includegraphics[width=.32\textwidth]{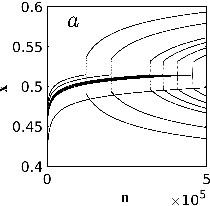} &
    \includegraphics[width=.31\textwidth]{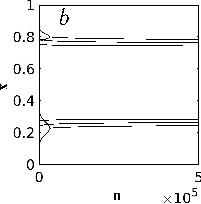} &
    \includegraphics[width=.31\textwidth]{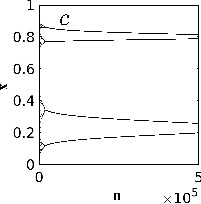} &   
  \end{tabular}
  \caption{Asymptotically period two trajectories for the fractional
difference Caputo logistic $\alpha$-family of maps Eqs.~(\ref{FalFacMap})~and~(\ref{LogEqG})  
with $\alpha=0.1$ and $K=2.5$: (a) seven trajectories with the initial 
conditions $x_0=0.0001$, $x_0=0.85+0.05i$, $i=0,1,2,3$, $x_0=0.12$, and 
$x_0=0.14$ (the leftmost bifurcation);   
(b) $x_0=0.6+0.1i$, $i=1,2,3$; (c) $x_0=0.9$. As $ n\rightarrow \infty $ 
all trajectories converge to the limiting values
$x_{lim1}=0.3045$ and  $x_{lim2}=0.7242$. 
The unstable fixed point is $x_{lim0}=(K-1)/K=0.6$.}
\label{fig7}
\end{figure}

Regular (without memory) models (systems of differential/difference
equations) may have stable and unstable solutions. It is not trivial 
for a model of a regular system to have a solution that converges to 
a stable state and then, after a while, breaks. On the other hand, as 
we mentioned before, cascade of bifurcations type solutions (CBTT) 
are essential features of discrete nonlinear fractional order (with 
power-law memory) systems.

In CBTT the system first converges to a 
low-periodicity (or fixed) point but then this solution becomes unstable, 
the system breaks and jumps into a higher periodicity solution. 
The simplest discrete nonlinear fractional/fractional difference model 
is the logistic map, which has power-law (or asymptotically power-law) 
memory and quadratic nonlinearity.   
In a wide range of parameters, when a fractional system asymptotically converges to a period $T > 1$ or chaotic trajectory, for a fixed set of parameters the evolution depends only on initial conditions. 
Figs.~6~and~7 show this dependence for the fractional and fractional difference Caputo logistic maps.
As one can see, for small values of initial conditions the system first slowly, in a regular way, starts converging to the unstable fixed point but then suddenly becomes unstable and bifurcates. The time of the regular convergence to the stable fixed point decreases with the increase in initial conditions. 
As it can be seen from Fig.~8, convergence to the stable fixed point follows the power law with the power $\alpha$ from
Eqs.~(\ref{FrCM})~and~(\ref{FalFacMap}). In the case of fractional difference maps, convergence to the power law is very clear (see Fig.~8c). In the case of fractional maps, for small values of $\alpha$ even convergence to the power law is slow (Fig.~8a), but Fig.~8b ($\alpha$=0.4) confirms that eventually the convergence follows the power law (for the power-law convergence of trajectories see also \cite{ME2,ME4,ME7,ME8,ME5}). 

\begin{figure}[htb]
\centering
  \begin{tabular}{@{}cccc@{}}
    \includegraphics[width=.31\textwidth]{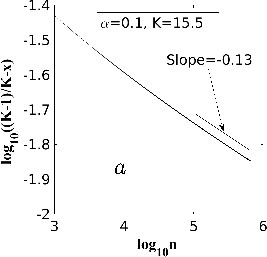} &
    \includegraphics[width=.31\textwidth]{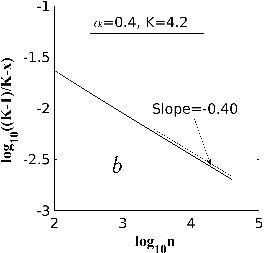} & 
    \includegraphics[width=.31\textwidth]{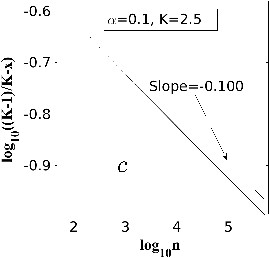} & 
  \end{tabular}
  \caption{Power-law convergence of trajectories to the fixed point (stable (b) or unstable (a) and (c)) for the fractional (a) and (b) and fractional difference (c) Caputo logistic $\alpha$-family of maps. Maps' parameters are indicated on the figures.}
\label{fig8}
\end{figure}

The lifespan in this model, the times of the regular convergence to a
stable fixed point until the bifurcation on a single trajectory (call 
it death, breaking of the stable fixed-point evolution, etc.), as can 
be seen from Fig.~9, is the exponential function of the initial condition
\begin{equation}
t=t_me^{-\gamma x},
\label{ExpT}
\end{equation}
where $t$ is the lifespan, $x$ is the initial condition and constants
$t_m$, which is the maximal possible age, and $\gamma$ depend on the map's 
parameters. 
\begin{figure}[htb]
\centering
  \begin{tabular}{@{}cccc@{}}
    \includegraphics[width=.49\textwidth]{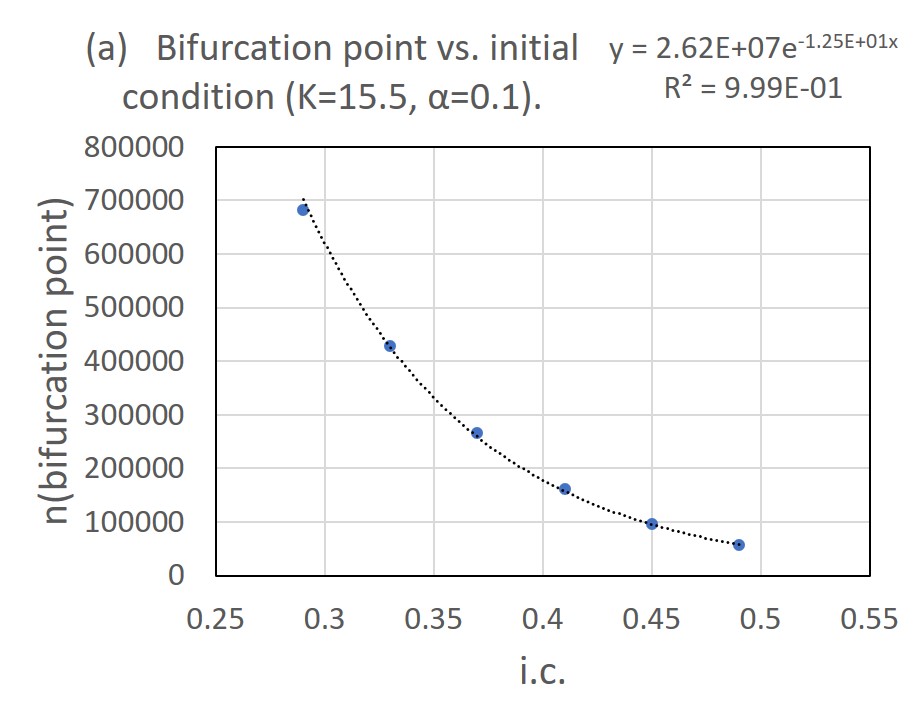} &
    \includegraphics[width=.49\textwidth]{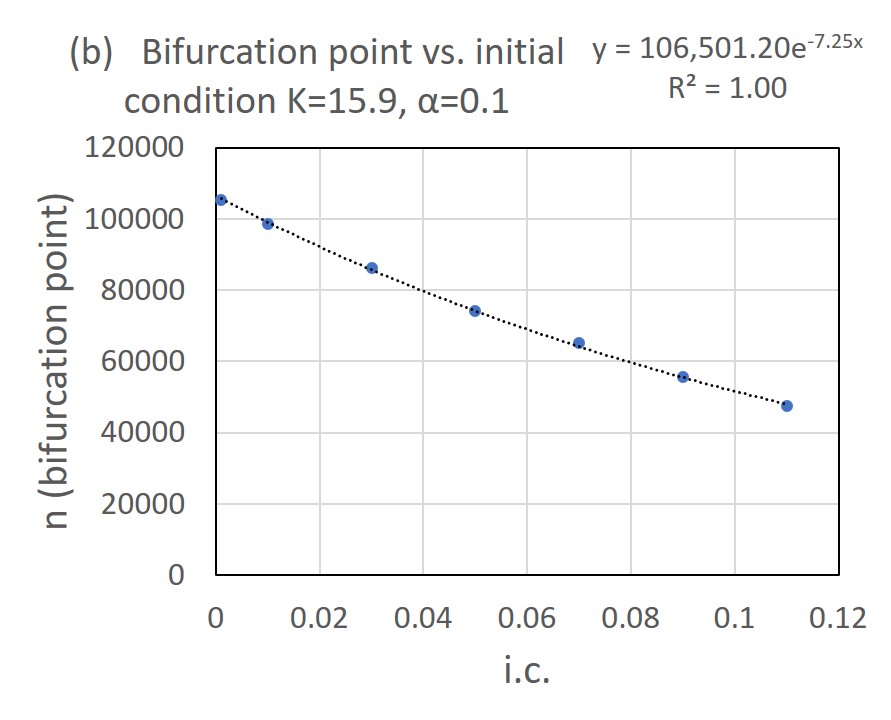} & \\ 
    \includegraphics[width=.49\textwidth]{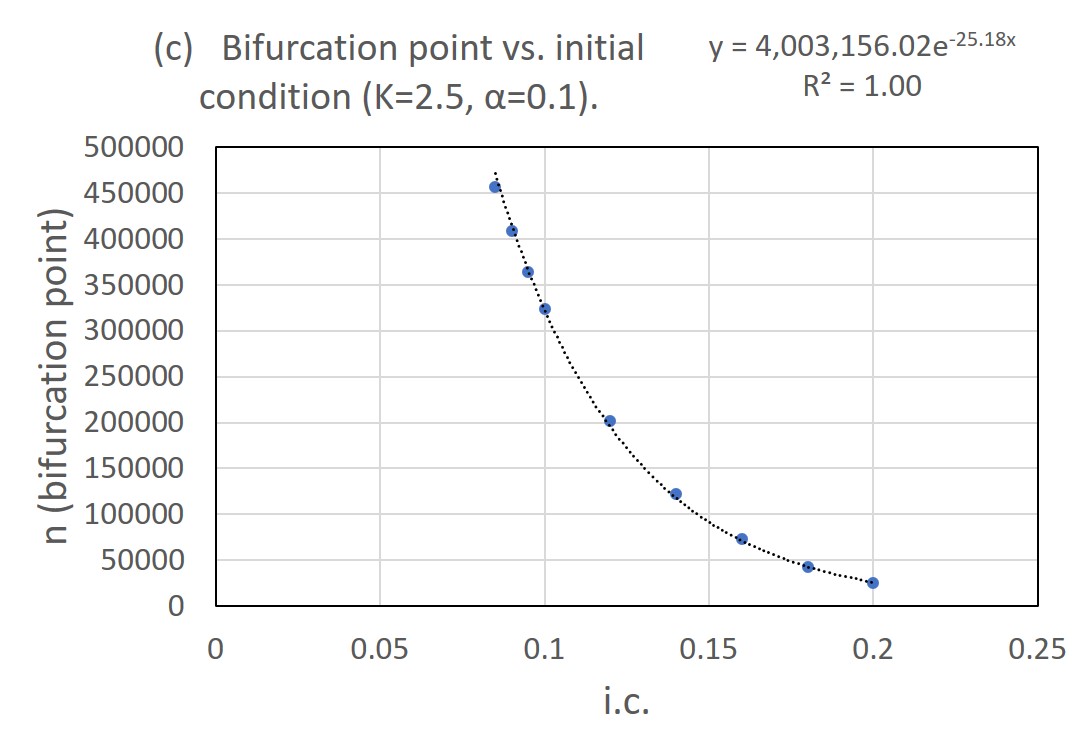} &
    \includegraphics[width=.49\textwidth]{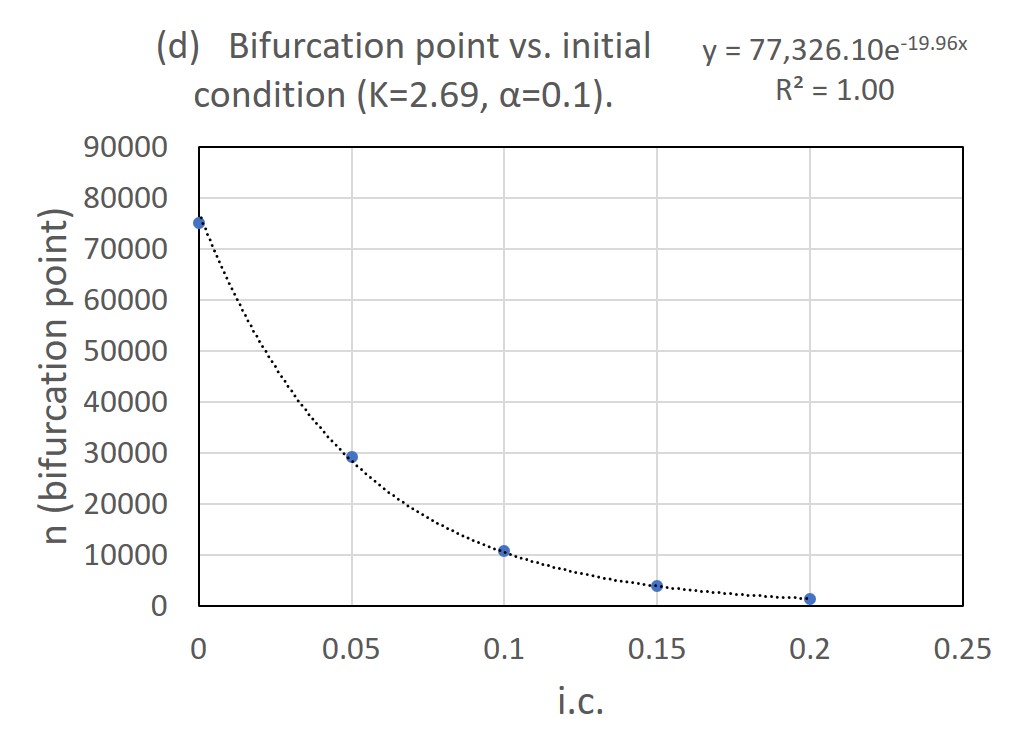} &
  \end{tabular}
  \caption{The time of the regular convergence to a stable fixed point until the bifurcation on a single trajectory for the fractional (a) and (b), and fractional difference (c) and (d) Caputo logistic $\alpha$-family of maps as a function of the initial condition. Maps' parameters are indicated on the figures.}
\label{fig9}
\end{figure}
If we assume the uniform distribution of the initial conditions, 
$dl(t)=dx(t)$, then calculating the force of mortality 
Eq.~(\ref{FM}), we'll obtain $\mu = 1/(t*(\ln{t_m}-\ln{t}))$. 
This assumption seems unrealistic and the corresponding formula 
for the force of mortality is very different from the Gompertz-Makeham law.

More realistic is to assume that the value of an initial condition (number 
of descendants produced at very early ages) is near zero and consider 
the evolution under random perturbations (mutations, interactions with 
the environment, etc.). There are various ways to introduce 
perturbations: they may be either uniformly distributed or normally; 
they may be added at each step of iterations starting either from 
the beginning of the evolution or at a certain age (this would correspond 
to a perfect cellular repair mechanism up to that age); they may be of 
a constant magnitude or increasing; and so on. 
In this publication we present only two cases: In Fig.~10 we present 
four different trajectories for the Caputo fractional logistic map with 
$\alpha=0.1$, $K=15.9$ and $x_0=0.0001$. In this case the maximal 
lifespan (without perturbations, this case is added to each subfigure) 
is approximately 106500 iterations (then the system bifurcates very 
fast). The perturbations $\sigma$ are uniformly distributed 
within the interval   $(-0.0005,0.0005)$ and are added at each 
step starting from $n=20001$. 
\begin{figure}[htb]
\centering
  \begin{tabular}{@{}cccc@{}}
    \includegraphics[width=.49\textwidth]{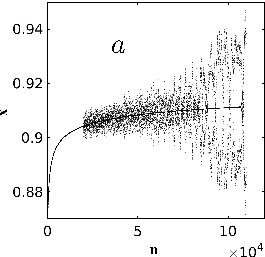} &
    \includegraphics[width=.49\textwidth]{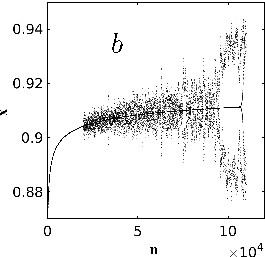} & \\ 
    \includegraphics[width=.49\textwidth]{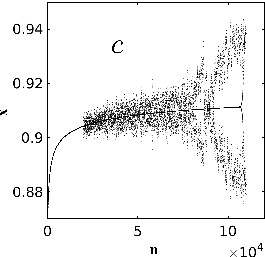} &
    \includegraphics[width=.49\textwidth]{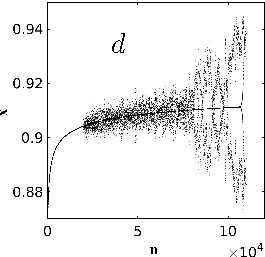} &
  \end{tabular}
  \caption{Four examples of the influence of perturbations in Caputo fractional logistic map with $\alpha=0.1$, $K=15.9$ and $x_0=0.0001$. Perturbations, uniformly distributed within the interval                $(-0.0005,0.0005)$, are added at each step of iterations starting from $n=20001$. The steady line, which breaks at the end, in the middle of each graph shows the unperturbed solution.}
\label{fig10}
\end{figure}
Fig.~11 presents four different trajectories for the fractional difference logistic map with $\alpha=0.1$, $K=2.69$ and $x_0=0.0001$.
In this case the maximal lifespan is approximately 77000 iterations. 
The perturbations $\sigma$ are normally distributed with the zero mean
value and the standard deviation 0.0005; perturbations are added at each
step starting from $n=16$.
\begin{figure}[htb]
\centering
  \begin{tabular}{@{}cccc@{}}
    \includegraphics[width=.49\textwidth]{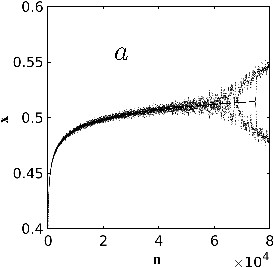} &
    \includegraphics[width=.49\textwidth]{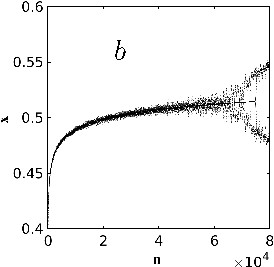} & \\ 
    \includegraphics[width=.49\textwidth]{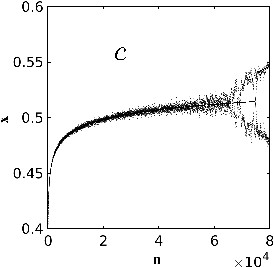} &
    \includegraphics[width=.49\textwidth]{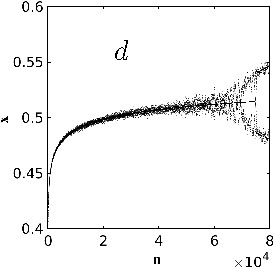} &
  \end{tabular}
  \caption{Four examples of the influence of perturbations in Caputo fractional difference logistic map with $\alpha=0.1$, $K=2.69$ and $x_0=0.0001$. Perturbations are normally distributed with the zero mean
value and the standard deviation 0.0005 and are added at each step of iterations starting from $n=16$. The steady line, which breaks at the end, in the middle of each graph shows the unperturbed solution.}
\label{fig11}
\end{figure}
\begin{figure}[!t]
\begin{center}
\includegraphics[width=0.77\textwidth]{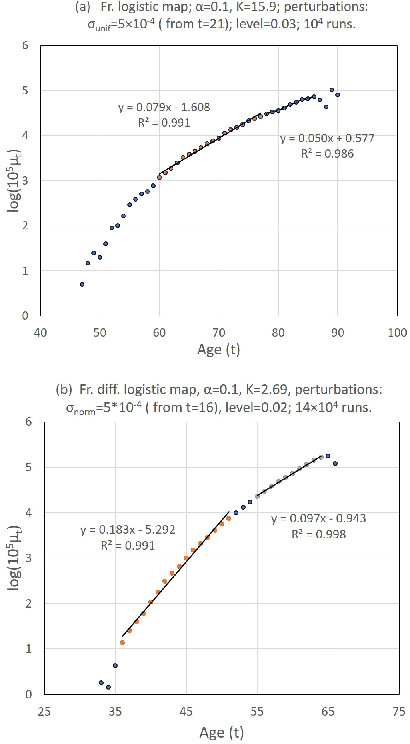}
\end{center}
\caption{The logarithm of the force of mortality as a function of age. All relevant parameters are indicted in the subfigures' titles. The maximal possible age (when there are no perturbations) in case (a) is 106
and in case (b) is 77.  
}
\label{fig12}
\end{figure}
These figures demonstrate the relative robustness of the solution with 
respect to perturbations. Still, in each case the bifurcation occurs 
much earlier compared to the unperturbed case.

To obtain the distribution of the times of the stable evolution prior to the bifurcation (lifespans) in the fractional/fractional difference under perturbation models, the author followed the standards described in Chapters 2~and~3 of the Gavrilovs' book \cite{GavGavBook}. In different cases (various maps, parameters, and perturbations) from tens of thousands to hundreds of thousands runs with various sets of random perturbations were performed. In each set of runs the magnitude of perturbations was much (hundreds or thousands times) smaller than the difference between two limiting values $x_{lim}$ of the asymptotically period two solution of the unperturbed equation. 
To distinguish the difference $|x_{n+1}-x_{n}|$ due to a bifurcation from the difference due to random perturbations, the value of the {\it{level}} (mentioned in Fig.~12), which is much larger than any reasonably probable value of a perturbation (e.g., 40 standard deviations in Fig.~12b), was selected. Then, the assumption was made that the steady power-law convergence to the fixed point terminates (death of an organism) when the difference $|x_{n+1}-x_{n}|$ for the first time exceeds the {\it{level}}.

Each run was terminated at the iteration $n_{lev}$ at which the magnitude of the difference between two consecutive values of $x$ for the first time exceeded the {\it{level}} and the value of $l_t$, where $t=[n_{lev}/1000]$ was increased by one. Then, the force of mortality $\mu_t$, Eq.~(\ref{FM}), was calculated according to the Eq.~(9) from the Chapter 2 of the book \cite{GavGavBook} as
\begin{equation}
\mu_t=\frac{1}{2}\ln\left(\frac{l_{t-1}}{l_{t+1}}\right).
\label{FMn}
\end{equation}
As in Chapter 3 of \cite{GavGavBook}, in this paper we draw the graphs of $\log(\mu_t\times10^5)$ as a function of the age $t$ and see if it fits a straight line (which implies the Gompertz-Makeham law).
 
Here (in Fig.~12) we present only results of the two sets of runs, but the Gompertz-Makeham law (with the various relative widths in time of nonzero changes in the number of survivors from the initial population $l_t$) was obtained in all cases. As one may see, the fit is approximately linear. It is not perfect, but it is not perfect in the graphs obtained from the real tables of mortality either.

\section{Conclusion}
\label{sec:5}

We are conceived and born programmed to develop in a certain way at a certain pace. If this program includes power-law memory, then we may be programmed to die. In some scenarios this death is inevitable, and the maximal possible lifespan may be calculated. The observed power-law in human memory and the viscoelastic nature of our organ tissues present reasonable arguments for modeling human individuals as fractional (with power-law memory) systems. Whether we are programmed to die or not, the power-law memory in the evolution of individual humans and separate organs may lead to the limit of the human lifespan. The same may be true for any living species. As one can see from the results of numerical simulations, the real lifespans may be significantly shorter than the maximal possible lifespan. In the table created for the example presented in Fig.~12b based on 140000 runs, when the maximal lifespan was 77, none survived at the age of 68; for the Fig.~12a the numbers are 10000 runs, maximal lifespan is 106, and none survived at the age of 94.      
Another significant fact is that in models with power-law memory the decrease at very large ages of the rate of increase of the force of mortality does not contradict the limited lifespan, which is impossible in memoryless models \cite{GavGavBook}.

If the assumption that living species develop as systems with power-law memory is correct, then the reasonable questions are where and how the power-law memory is recorded in our DNA and whether we may manage and correct this recording. The author understands that his proposition raises more questions than it produces answers and welcomes any discussions on the topic of the power law in human evolution. 

With regards to the theory of stability of fractional (with power-law memory) systems, the simulations presented in this paper demonstrate the robustness of the factional systems with regards to various types of perturbations. As can be seen from Figs.~10~and~11, during a significant interval of time in the evolution of a fractional system, memory may prevail in competition with perturbations and preserve the unperturbed evolution.

\begin{acknowledgement}  

The author acknowledges support from the Joseph
Alexander Foundation, Yeshiva University. The author expresses his gratitude to Prof. Sylvain E. Cappell for the support at the Courant Institute where the computer simulations were performed. The author also expresses his gratitude to Virginia Donnelly for technical help and valuable discussions.

\end{acknowledgement}

\end{document}